\documentclass[aps, prd, superscriptaddress, twocolumn]{revtex4-1}

\usepackage{wallpaper}
\usepackage{bm}
\usepackage{amsmath}
\usepackage[colorlinks,linkcolor=blue,anchorcolor=blue,citecolor=red]{hyperref}

\begin{document}

\title{Strangelets at finite temperature}

\author{Hao-Song~You}
%\email{213303128@stu.yzu.edu.cn}
\affiliation{Center for Gravitation and Cosmology, College of Physical Science and Technology, Yangzhou University, Yangzhou 225009, China}

\author{Huai-Min~Chen}
%\email{chenhuaimin@wuyiu.edu.cn}
\affiliation{School of Mechanical and Electrical Engineering, Wuyi University, Wuyishan 354300, China}

\author{Jian-Feng~Xu}
%\email{jfxu@aynu.edu.cn}
\affiliation{College of information Engineering, Suqian University, Suqian 223800, China}

\author{Cheng-Jun~Xia}
\email{cjxia@yzu.edu.cn}
\affiliation{Center for Gravitation and Cosmology, College of Physical Science and Technology, Yangzhou University, Yangzhou 225009, China}

\author{Ren-Xin~Xu}
%\email{r.x.xu@pku.edu.cn}
\affiliation{School of Physics, Peking University, Beijing 100871, China}
\affiliation{Kavli Institute for Astronomy and Astrophysics, Peking University, Beijing 100871, China}

\author{Guang-Xiong~Peng}
%\email{gxpeng@ucas.ac.cn}
\affiliation{School of Nuclear Science and Technology, University of Chinese Academy of Sciences, Beijing 100049, China}
\affiliation{Theoretical Physics Center for Science Facilities, Institute of High Energy Physics, Beijing 100049, China}
\affiliation{Synergetic Innovation Center for Quantum Effects and Application, Hunan Normal University, Changsha 410081, China}

\date{\today}

\begin{abstract}
We study the properties of strangelets at finite temperature $T$, employing an equivparticle model that incorporates both linear confinement and leading-order perturbative interactions with density-dependent quark masses. The shell effects are analyzed by solving the Dirac equations for quarks within the mean-field approximation. As temperature increases, these effects weaken due to the occupation probability of single-particle levels being governed by the Fermi-Dirac statistics, a phenomenon known as shell dampening. Surprisingly, the surface tension, derived from a liquid-drop formula, does not decrease with temperature but instead rises until it peaks at $T \approx 20-40$ MeV. At this temperature, shell corrections become negligible, and the formula provides a reasonable approximation for the free energy per baryon of strangelets. However, the curvature term decreases with $T$ despite the presence of shell effects. The neutron and proton emission rates are determined microscopically by the external nucleon gas densities that are in equilibrium with strangelets. These emission rate generally increases with $T$ for stable strangelets, but decrease for those that are unstable to nucleon emission at $T$ = 0. The other properties of $\beta$-stable strangelets obtained with various parameter sets are presented as well. The results indicated in this work are useful for understanding the products of binary compact star mergers and heavy-ion collisions.

\end{abstract}

%\pacs{21.65.Qr, 12.39.-x, 25.75.Nq}
%21.65.Qr Quark matter
%12.39.-x Phenomenological quark models
%25.75.Nq Quark deconfinement, quark-gluon plasma production, and phase transitions.

\maketitle

\section{\label{sec:intro}Introduction}

Quantum Chromodynamics (QCD) predicts the existence of multiquark states beyond mesons and baryons, in which the number of quarks exceeds three. As the density of nuclear matter increases, a phase transition from confine hadronic matter to deconfined quark matter is expected~\cite{Fukushima2005_PRD71-034002}. In 1971, Bodmer suggested the possibility of ``collapsed nuclei" with a mass number $A$ exceeding 1 and a significantly reduced radius when compared to typical nuclei~\cite{Bodmer1971_PRD4-1601}. Subsequently, Witten posited that strange quark matter (SQM), containing approximately equal numbers of u, d, and s quarks along with a small quantity of electrons, could constitute the ground state of strongly interacting systems~\cite{Witten1984_PRD30-272}. Based on various quark models such as the bag model,  it has been demonstrated that strange quark matter (SQM) is more stable than nuclear matter in the extensive parameter space~\cite{Terazaw1989_JPSJ58-3555}. The absolute stability of SQM allows for the possible existence of strangelets~\cite{Farhi1984_PRD30-2379,Berger1987_PRC35-213,Gilson1993_PRL71-332,Peng2006_PLB633-314}, nuclearites~\cite{Rujula1984_Nature312-734,Lowder1991_NPB24-177}, meteor-like compact ultradense objects~\cite{Rafelski2013_PRL110-111102}, and strange stars~\cite{Itoh1970_PTP44-291,Alcock1986_ApJ310-261,Haensel1986_AA160-121}. Despite this, when the effect of dynamical chiral symmetry breaking is taken into account, SQM may become unstable~\cite{Buballa1999_PLB457-261,Klahn2015_ApJ810-134}, and non-strange quark matter ($ud$QM) may constitute the true ground state~\cite{Holdom2018_Phys.Rev.Lett.120-222001}. Moreover, instead of being in a deconfined state, a stable solid phase made up of strangeons (clusters of quarks with three-light-flavor symmetry) may prevail, which permit the existence of minuscule strangeon nuggets throughout our universe~\cite{Xu2003_ApJ596-L59,Miao2022_IJMPE0-2250037,Xu2019_AIPCP2127-020014}.
These exotic objects are thought to originate from various sources, including heavy-ion collisions~\cite{Weiner2006_IJMPE15-37}, the mergers of binary compact stars~\cite{Madsen2002_JPG28-1737,Madsen2005_PRD71-014026,Bauswein2009_PRL103-011101,Lai2018_RAA18-024,Lai2021_RAA21-250}, type II supernova explosions~\cite{Vucetich1998_PRD57-5959}, and the hadronization process in the early universe~\cite{Witten1984_PRD30-272}. Substantial endeavors have been invested in the quest for these enigmatic objects, yet conclusive evidence remains elusive~\cite{Finch2006_JPG32-S251,Burdin2015_PR582-1}. Consequently, it is imperative for us to unravel the intricacies of their properties to ultimately validate or refute their very existence.

Given its non-perturbative nature and the notorious sign problem in lattice simulations, at present we must rely on various QCD-inspired effective models to reveal the properties of these objects. For instance, utilizing the MIT bag model with bag constant $B$, where the surface tension $\sigma^{1/3}\approx B^{1/4}$, Farhi and Jaffe discovered that the surface tension plays a significant role in the stability of strangelets~\cite{Farhi1984_PRD30-2379}. Later on, Berger and Jaffe proposed a mass formula for strangelets and investigated their possible decay channels~\cite{Berger1987_PRC35-213}. Small strangelets are almost uniformly charged, while the charge screening effects start to play a role for larger strangelets~\cite{Heiselberg1993_PRD48-1418}. By considering a reasonable surface tension and incorporating Coulomb interactions, it was found that large strangelets are likely to be stable against fission~\cite{Heiselberg1993_PRD48-1418}. Furthermore, if the surface tension $\sigma$ is sufficiently small, there exist strangelets of a certain size that are more stable than others~\cite{Alford2006_PRD73-114016}. In such cases, the surfaces of strange stars may fragment into crystalline crusts composed of strangelets or $ud$QM nuggets and electrons~\cite{Xia2022_PRD106-034016,Jaikumar2006_PRL96-041101}, or even form low-mass, large-radius strangelet dwarfs~\cite{Alford2012_JPG39-065201}. It has been demonstrated for small strangelets and $ud$QM nuggets that the curvature contribution is significant~\cite{Lugones2021_PRC103-035813}, leading to the development of the multiple reflection expansion method (MRE) to analyze surface and curvature corrections analytically~\cite{Madsen1993_PRL70-391,Madsen1993_PRD47-5156,Madsen1994_PRD50-3328}. The shell effects in strangelets were also examined within the framework of the MIT bag model, which significantly affects the properties of small strangelets~\cite{Madsen1994_PRD50-3328,Terazaw1989_JPSJ58-3555,Terazaw1989_JPSJ58-4388,Terazaw1990_JPSJ59-1199,Oertel2008_PRD77-074015}.

The studies on strangelets and $ud$QM nuggets are typically conducted at zero temperature, but their creation and survival often occur at large temperatures. Therefore, it is necessary to investigate the properties of these objects at finite temperatures. In the framework of MIT bag model, the properties of strangelets at finite temperatures were investigated, where clear shell structures persist up to about $T$ = 10 MeV were identified~\cite{Mustafa1996_PRD53_5136--5141,Mustafa1997_PRC56-420}. The cosmological quark-hadron transition reveals a significant suppression of evaporation and boiling processes for quark nuggets in a color-flavor locked state~\cite{Lugones2004_PRD69-063509}. However, it is noteworthy that the MIT bag model assumes an infinitely large boundary, whereas lattice QCD suggests linear confinement for quarks, resulting in distinct surface density profiles and consequently impacting various properties of strangelets as indicated in our previous study~\cite{Xia2018_PRD98-034031}. We thus employ an equivparticle model in this study to investigate the properties of strangelets, wherein both the linear confinement and leading-order perturbative interactions are incorporated by considering density-dependent quark masses~\cite{Peng2000_PRC62-025801,Wen2005_PRC72-015204,Wen2007_JPG34-1697,Chen2012_CPC36-947,Xia2014_PRD89-105027}. Note that the properties of strangelets at finite temperatures have been examined in the framework of the equivparticle model, where the interface effects are treated using the MRE method~\cite{Wen2005_PRC72-015204,Zhang2003_PRC67_015202,Chen2022_PRD105-014011}. The results indicate that the mass, radius, and strangeness per baryon increase with temperature, while the charge-to-mass ratio decreases with temperature~\cite{Chen2022_PRD105-014011}.

The paper is organized as follows. In Sec.\ref{Sec.theoretical}, we present the theoretical framework of the equivparticle model, where the Lagrange density of equivparticle model with density dependent quark masses is given in Sec.\ref{sec:the_Lagrangian}, the mean-field approximation (MFA) is introduced in Sec.\ref{Sec:Strangelet} with the quark wavefunctions obtained by solving Dirac equations, and the emission rates of neutrons and protons can be determined based on the densities of external neutron and proton gases that are in equilibrium with the strangelet~\cite{Zhu2014_PRC90-054316} in Sec.\ref{Sec:emi}. Our numerical results on the properties strangelets at finite temperature are presented in Sec\ref{sec:Results}. Finally, we draw our conclusion in Sec.\ref{sec:summary}.

\section{\label{Sec.theoretical} Theoretical framework}
\subsection{\label{sec:the_Lagrangian}Lagrangian density}
The Lagrangian density of the equivparticle model can be given as
\begin{equation}
\mathcal{L} =  \sum_{i=u,d,s} \bar{\Psi}_i \left[ i \gamma^\mu \partial_\mu - m_i(n_\mathrm{b}) - e q_i \gamma^\mu A_\mu \right]\Psi_i
             - \frac{1}{4} A_{\mu\nu}A^{\mu\nu},  \label{eq:Lgrg_all}
\end{equation}
where $\psi_i$ is the Dirac spinor of quark flavor $i$, $m_i$ the mass, $q_i$ the charge ($q_u = 2e/3$ and $q_d = q_s = e/3$), and $A_\mu$ the photon field with the field tensor
\begin{equation}
A_{\mu\nu} = \partial_\mu A_\nu - \partial_\nu A_\mu.
\end{equation}
In the equivparticle model, the strong interactions are considered with density and temperature-dependent quark masses and quarks are treated as quasi-free particles. The mass of quark $i$ is fixed as
\begin{equation}
  m_i(n_{\mathrm b})=m_{i0} + m_\mathrm{I}(\{n_j\},T)\label{Eq:mnbC}
\end{equation}
Here $n_j = \left< \bar{\psi}_j \gamma^0 \psi_j \right>$ represents the number density for quark flavor $j$, $T$ the temperature, and $m_{u0}=2.2$ MeV, $m_{d0}=4.4$ MeV, $m_{s0}=96$ MeV the current quark masses~\cite{PDG2016_CPC40-100001}. Considering the quark confinement effect in bag model~\cite{Fowler1981_ZPC9-271,Chakrabarty1989_PLB229-112,Benvenuto1995_PRD51-1989}, the density dependent quark masses for zero temperature can be given by
\begin{equation}
  m_\mathrm{I}(n_{\mathrm b})=\frac{B}{3 n_{\mathrm b}} \label{Eq:MITbag}
\end{equation}
where $n_{\mathrm b} = \sum_{i=u,d,s} n_i/3$ is the baryon number density and $B$ is the bag constant. Alternatively, to reflect the contributions of linear confinement and in-medium chiral condensates, a cubic-root scaling~\cite{Peng1999_PRC61-015201} was proposed
\begin{equation}
  m_\mathrm{I}(n_{\mathrm b})= Dn_{\mathrm b}^{-1/3} \label{Eq:D}
\end{equation}
where $D=-3(2/\pi)^{1/3}\sigma n^*/\sum_q\left<\bar{q}q\right>_0$ represents the confinement strength, which is connected to the
string tension of linear confinement and vacuum chiral condensates.
In order to take into account both the one-gluon-exchange interaction and the leading-order perturbative interactions~\cite{Chen2012_CPC36-947,Xia2014_PRD89-105027}, Eq.~(\ref{Eq:D}) is modified as follows
\begin{equation}
  m_\mathrm{I}(n_{\mathrm b})= Dn_{\mathrm b}^{-1/3} + Cn_{\mathrm b}^{1/3} \label{Eq:D}
\end{equation}
where the perturbative strength parameter $C \approx \pi^{2/3}\sqrt{{2\alpha_s}/{3\pi}}$, with the strong coupling constant $\alpha_s$. Depending on the sign of $C$, the second term of Eq.~(\ref{Eq:D}) corresponds to the contribution from one-gluon-exchange interaction ($C<0$)~\cite{Chen2012_CPC36-947} or leading-order perturbative interaction ($C>0$)~\cite{Xia2014_PRD89-105027}. To adapt to the phase transition of deconfinement at large temperatures, this mass scaling was subsequently extended to be temperature-dependent, i.e.,
\begin{equation}
  m_\mathrm{I}(n_{\mathrm b})= Dn_{\mathrm b}^{-1/3}(1+\frac{8T}{\Lambda}\exp^{-\frac{\Lambda}{T}})^{-1} + C n_{\mathrm b}^{1/3}(1+\frac{8T}{\Lambda}\exp^{-\frac{\Lambda}{T}}) \label{T-depend}
\end{equation}
where $\Lambda = 280$  MeV is a temperature scale parameter corresponding to the critical temperature $T_c \approx 175$ MeV~\cite{Wen2005_PRC72-015204}.
Given that strangelets do not exist at high temperatures above $T_c$ and the variation of $m_{I}$ with respect to temperature is generally small below $T_c$, we have chosen to neglect the temperature dependence of quark masses and instead utilize Eq.~(\ref{Eq:D}) in this study.

\subsection{\label{Sec:Strangelet} Strangelets in MFA} 
For spherically symmetric strangelets, the Dirac spinor of quarks can be expanded as
\begin{equation}
 \psi_{n\kappa m}({\bm r}) =\frac{1}{r}
 \left(\begin{array}{c}
   iG_{n\kappa}(r) \\
    F_{n\kappa}(r) {\bm\sigma}\cdot{\hat{\bm r}} \\
 \end{array}\right) Y_{jm}^l(\theta,\phi)\:,
\label{EQ:RWF}
\end{equation}
where $G_{n\kappa}(r)/r$ and $F_{n\kappa}(r)/r$ are the radial wave functions for upper and lower components, while $Y_{jm}^l(\theta,\phi)$ is the spinor spherical harmonics. The quantum number $\kappa$ is defined by the angular momenta ($l,j$) as $\kappa = (-1)^{j+l+1/2}(j+1/2)$ with $j=l\pm1/2$.

By utilizing the mean-field approximation and inserting Eq.~(\ref{EQ:RWF}) into the Dirac equation, we can easily acquire the one-dimensional radial Dirac equation through the integration of the angular component, i.e.,
\begin{equation}
 \left(\begin{array}{cc}
  V_{iV} + V_{iS}                                                   & {\displaystyle -\frac{\mbox{d}}{\mbox{d}r} + \frac{\kappa}{r}}\\
  {\displaystyle \frac{\mbox{d}}{\mbox{d}r}+\frac{\kappa}{r}} & V_{iV} - V_{iS}                       \\
 \end{array}\right)
 \left(\begin{array}{c}
  G_{in\kappa} \\
  F_{in\kappa} \\
 \end{array}\right)
 = \varepsilon_{in\kappa}
 \left(\begin{array}{c}
  G_{in\kappa} \\
  F_{in\kappa} \\
 \end{array}\right) \:,
\label{Eq:RDirac}
\end{equation}
where $\varepsilon_{in\kappa}$ is the single particle energy. The mean-field scalar and vector potentials of quarks can be obtained as
\begin{eqnarray}
 V_{iS} &=& m_{i0} +
 m_\mathrm{I}(n_\mathrm{b}), \label{Eq:Vs}\\
 V_{iV} &=& \frac{1}{3}\frac{\mbox{d} m_\mathrm{I}}{\mbox{d} n_\mathrm{b}}\sum_{i=u,d,s}  n_i^\mathrm{s} + e q_i A_0, \label{Eq:Vv} \label{Eq:Vv}
\end{eqnarray}
where there exist common terms of the scalar and vector potentials $V_{S} = m_\mathrm{I}(n_\mathrm{b})$ and $V_{V} = \frac{1}{3}\frac{\mbox{d} m_\mathrm{I}}{\mbox{d} n_\mathrm{b}}\sum_{i=u,d,s}  n_i^\mathrm{s}$.
In addition, it should be noted that the scalar potential in Eq.~(\ref{Eq:Vs}) now incorporates the current masses of quarks. The vector potential arises as a result of the density-dependence of quark masses and is crucial for ensuring thermodynamic self-consistency. The Klein-Gordon equation for photons is given by
\begin{equation}
- \nabla^2 A_0 = e n_\mathrm{ch}. \label{Eq:K-G}
\end{equation}
where $n_\mathrm{ch}=\sum_iq_in_i$ is the charge density $q_u=2/3$, $q_d=-1/3$, and $q_s=-1/3$.

Once the single particle energies for quarks are established, their occupation probability is subsequently determined using the Fermi-Dirac distribution.
\begin{equation}
f_{in\kappa} = [1+e^{(\varepsilon_{in\kappa}-\mu_i)/T}]^{-1}. \label{Eq:F-D}
\end{equation}
where $\mu_i$ represents the chemical potential of quark favor $i$, and we focus solely on the $\beta$-equilibrated case with
\begin{equation}
\mu_u=\mu_d=\mu{_s}=\mu_\mathrm{b}/3 \label{mub}.
\end{equation}
We have adopted the no-sea approximation and disregarded any contributions from anti-quarks, which is justified at the small temperatures considered here. As neutrinos can freely escape the system and electrons have minimal impact on the properties of strangelets with radii $R \gtrsim 40$ fm, the contributions of neutrinos and electrons are disregarded by assuming their chemical potentials to be zero.  The scalar and vector densities are fixed by
\begin{subequations}
\begin{eqnarray}
 n_i^\mathrm{s}(r) &=& \sum_{n,\kappa} \frac{3|\kappa|f_{in\kappa}}{2\pi r^2}
 \left[|G_{in\kappa}(r)|^2-|F_{in\kappa}(r)|^2\right] \:,
\\
 n_i(r) &=& \sum_{n,\kappa} \frac{3|\kappa|f_{in\kappa}}{2\pi r^2}
 \left[|G_{in\kappa}(r)|^2+|F_{in\kappa}(r)|^2\right] \:,
\end{eqnarray}%
\label{Eq:Density}%
\end{subequations}%
where the degeneracy factor $3(2j+1)=6|\kappa|$ of each single particle levels is considered. The quark numbers $N_i=\int 4\pi r^2 n_i(r)dr$ $(i=u,d,s)$ are obtained by integrating the density $n_i(r)$ in coordinate space.

Finally, the total mass and free energy of a strangelet can be obtained with
\begin{eqnarray}
M &=& \sum_{i,n,\kappa} 6|\kappa|f_{in\kappa}\varepsilon_{in\kappa} - \int 12\pi r^2 n_\mathrm{b}(r) V_V(r) \mbox{d}r \label{Eq:M} \\
  &\mathrm{}& - \int 2\pi r^2 n_\mathrm{ch}(r) e A_0(r) \mbox{d}r, \nonumber    \\
F &=& \sum_{i,n,\kappa} 6|\kappa|f_{in\kappa}  \left\{ \mu_i -T \ln  \left[1+e^{(-\varepsilon_{in\kappa}-\mu_{i})/T}\right] \right\} \label{Eq:Free energy} \\
  &\mathrm{}&  - \int 12\pi r^2 n_\mathrm{b}(r) V_V(r) \mbox{d}r   - \int 2\pi r^2 n_\mathrm{ch}(r) e A_0(r) \mbox{d}r. \nonumber
\end{eqnarray}
The last term of mass and free energy represents the contribution of the Coulomb energy. %In addition, the entropy is determined based on the fundamental thermodynamic relation, i.e.,
%\begin{equation}
%  S = \frac{M-F}{T} \label{Eq:entropy}
%\end{equation}

For a fixed baryon number $A$, temperature $T$, and parameter set ($C,D$), we solve the Dirac Eq.~(\ref{Eq:RDirac}), mean field potential Eqs.~(\ref{Eq:Vs}) and (\ref{Eq:Vv}), and density Eq.~(\ref{Eq:Density}) inside a box by iteration in coordinate space adopting the grid width less than 0.005 fm. Note that in our calculation we have introduced cutoffs on the quantum numbers  $n\leq n_{max}$ and $|\kappa|\leq\kappa_{max}$, which are fixed by the criterion $6|\kappa|f_{in\kappa}\leq10^{-6}$.

\subsection{\label{Sec:emi}Nucleon emission rates}
Having fixed the strangelet's properties in Sec.~\ref{Sec:Strangelet}, the neutron and proton emission widths can be estimated by examining the external neutron and proton gas densities that are in equilibrium with the strangelet. These densities are obtained using the standard formulae for nuclear reaction rates in nucleosynthesis theory~\cite{Zhu2014_PRC90-054316}.i.e.,
\begin{equation}
  \Gamma_{p,n}=n_{p,n}\left<\sigma_{n,p}v_{n,p}\right> \label{Eq:emi}
\end{equation}
where $\sigma_{n,p}$ is the neutron and proton cross sections of the strangelet, while $v_{n,p}$ is the velocities of protons and  neutrons $\left<\sigma_{n,p}v_{n,p}\right>$ represent statistical average over the states in the external gas. In addition, we can also derive the particle number densities of neutron and proton gases outside the hot strangelet, i.e.,
\begin{equation}
  n_{p,n}=\frac{1}{\pi^2}\int_0^{\infty}\left[ 1+e^{\left(\sqrt{p^2+m_{p,n}^2}-\mu_\mathrm{b}\right)/T}\right]^{-1} p^{2}\mathrm{d}p. \label{Eq:N-dens}
\end{equation}
Here the chemical potential $\mu_\mathrm{b}$ is fixed by Eq.~(\ref{mub}) where the strangelet is in thermodynamic equilibrium with the nucleon gas. It is worth mentioning that due to the minimal impact of antiparticles at small temperatures, their contribution is not considered in this work.

The statistical average $\left<\sigma_{n,p}v_{n,p}\right>$ can be calculated with
\begin{equation}
\left<\sigma_{i}v_{i}\right>=\frac{\int_0^{\infty}\sigma_i(\varepsilon_i)f(\varepsilon_i)v_i(\varepsilon_i)\sqrt{\varepsilon_i}\mathrm{d}\varepsilon_i}{\int_0^{\infty}f({\varepsilon_i})\sqrt{\varepsilon_i}\mathrm{d}\varepsilon_i}, \label{Eq:s-a}
\end{equation}
where $f(\varepsilon_i)$ represents the Fermi-Dirac distribution in Eq.~(\ref{Eq:F-D}) with the discretized single particle energy $\varepsilon_{in\kappa}$ replaced by the cintinuum one $\varepsilon_i$ and chemical potential $\mu_i$ by $\mu_\mathrm{b}-m_i$. The kinetic energy is connected to the velocity $v_i$ of the nucleons, i.e.,
\begin{subequations}
\begin{eqnarray}
 \varepsilon_i &=&\sqrt{p_i^2+m_i^2}-m_i \approx m_i^2v_i^2/2  \:,\\
 v_i(\varepsilon_i) &=& \sqrt{\frac{2\varepsilon_i}{m_i}} \:,
\end{eqnarray}%
\label{Eq:k-energy}%
\end{subequations}%
As the neutron is electrically neutral, we can straightforwardly utilize the geometrical cross section.i.e.,
\begin{equation}
\sigma_n = \pi R^2
\end{equation}
where $R$ is the radius of the corresponding strangelet and we take its value at vanishing quark densities. When calculating the capture cross section for protons, it is essential to account for the Coulomb interaction~\cite{Lai2021_RAA21-250,Wong1973_PRL31-766}. Here we utilize the Hill-Wheeler formula~\cite{Hill1953_PR089-1102} and assume a typical Coulomb barrier width of $\omega_0$ = 4 MeV for nuclear reactions, i.e.,
\begin{equation}
\sigma_p(\varepsilon_p)=\frac{R^2\omega_0}{2\varepsilon_p}\ln\left\{ 1+\mathrm{exp}\left[\frac{2\pi(\varepsilon_p-\varepsilon_C)}{\omega_0}\right]\right\}. \label{Coul-section}
\end{equation}
Here the Coulomb barrier height is determined by the box boundary with $\varepsilon_C=\alpha Z/R$, where $\alpha=1/137.036$ is the fine structure constant.
%%%%%%%%%%%%%%%%%%%%%%%%%%%%%%%%%%%%%%%%%%%%%%%%%%%%%%%%%%

\section{\label{sec:Results} Results and discussions}
For a strangelet with given total baryon number $A$, temperature $T$, and the parameter set ($C,\sqrt{D}$), the free energy $F$ in Eq.~(\ref{Eq:Free energy}) reaches minimum, i.e., fulfilling the $\beta$-stable condition. The density profiles can also be obtained with Eq.~(\ref{Eq:Density}). The other properties of strangelet can be derived with the formulae presented in Sec.~\ref{Sec.theoretical}. The neutron and proton emission widths are then estimated by Eq.~(\ref{Eq:emi}). Previously, we have conducted investigations to understand the properties of strangelets and $ud$QM nuggets at temperatures close to absolute zero. These properties are greatly influenced by the strengths of the confinement interaction $D$ and perturbation interaction $C$. In this investigation, we are particularly interested in exploring the effects of finite temperature on the characteristics of strangelets.

\begin{figure}
\includegraphics[width=\linewidth]{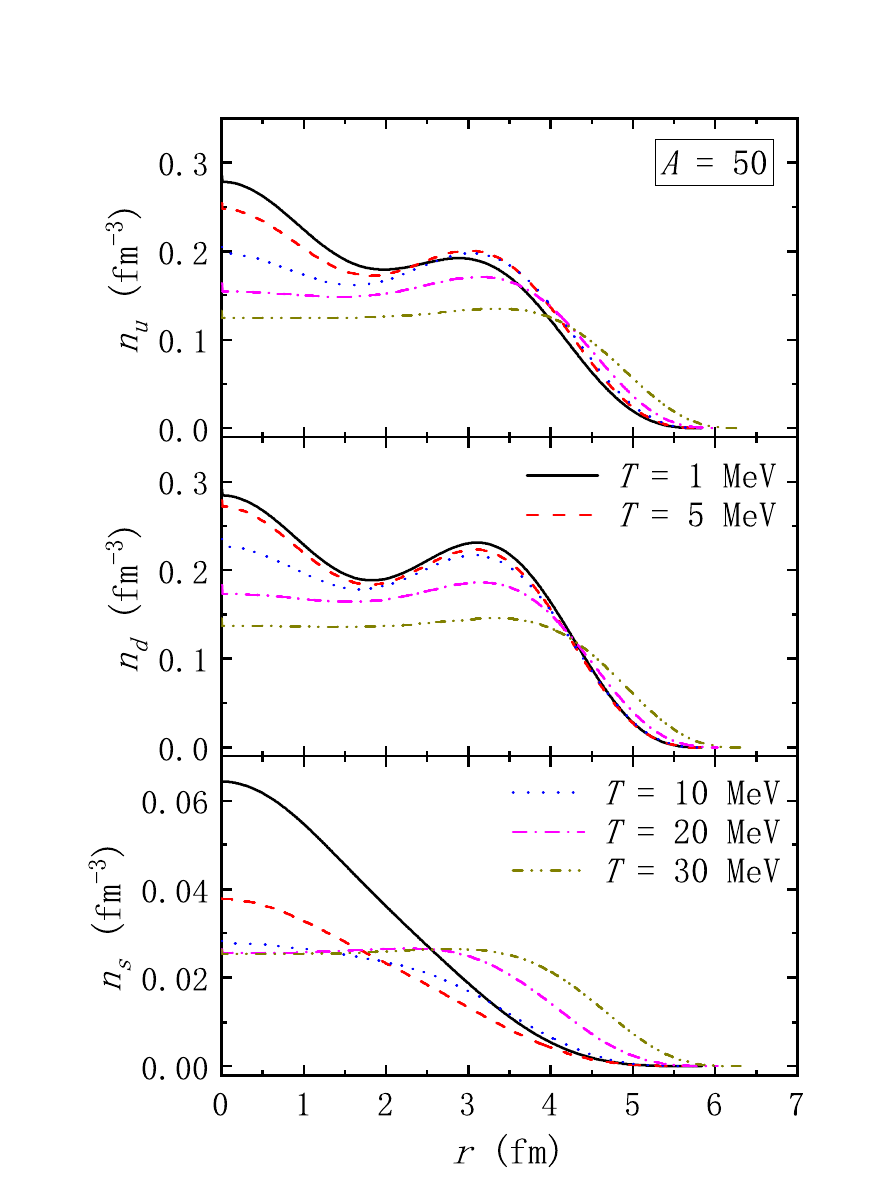}
\caption{\label{Fig:udsDens.pdf} Density profiles of $u$-, $d$-, $s$-quarks for strangelets at baryon number $A$ = 50 with various temperatures, where the parameter set $C$ = 0.4 and $D$ = 129 MeV is adopted.}
\end{figure}

\begin{figure}
\includegraphics[width=\linewidth]{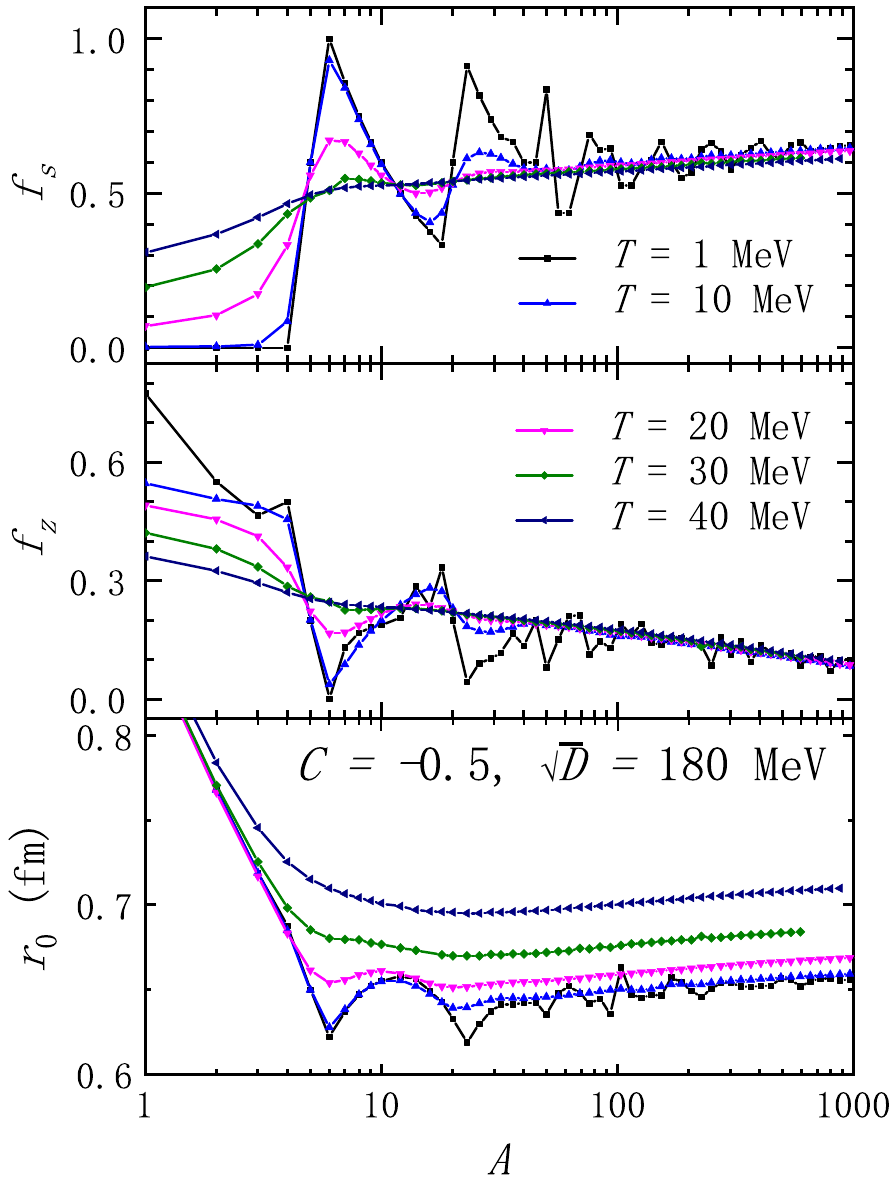}
\caption{\label{Fig:fsfzr0.pdf} Charge-to-mass ratio $f_z$ strangeness per baryon $f_s$,
and ratio of root-mean-square radius to baryon number $r_0$ for $\beta$-stable strangelets obtained at various temperatures.}
\end{figure}

In Fig.~\ref{Fig:udsDens.pdf}, we present the density profiles of $u$, $d$, and $s$ quarks for strangelets with a baryon number of A = 50 adopting temperatures ranging from $T$ = 1 to 30 MeV. These profiles were obtained using the parameter set $C$ = 0.4 and $\sqrt{D}$ = 129 MeV. As the temperature rises, the densities of $u$, $d$ quarks experience significant decreases at temperatures exceeding 20 MeV. However, at temperatures below 20 MeV, the shell effects become crucial, resulting in an increase in the densities of $u$ and $d$ quarks while the density of $s$ quarks decreases. This is primarily due to the decreasing strangeness per baryon, $f_s = N_s/A$, of the $\beta$-stable strangelet as the temperature rises, as demonstrated in Fig.~\ref{Fig:fsfzr0.pdf}. The density distribution near the quark-vacuum interface becomes more diffused with increasing temperature. The variations in the density profiles of $u$ and $d$ quarks generally exhibit similar trends as temperature increases, while the density of $s$ quarks shows more significant variation.

\begin{figure}
\includegraphics[width=\linewidth]{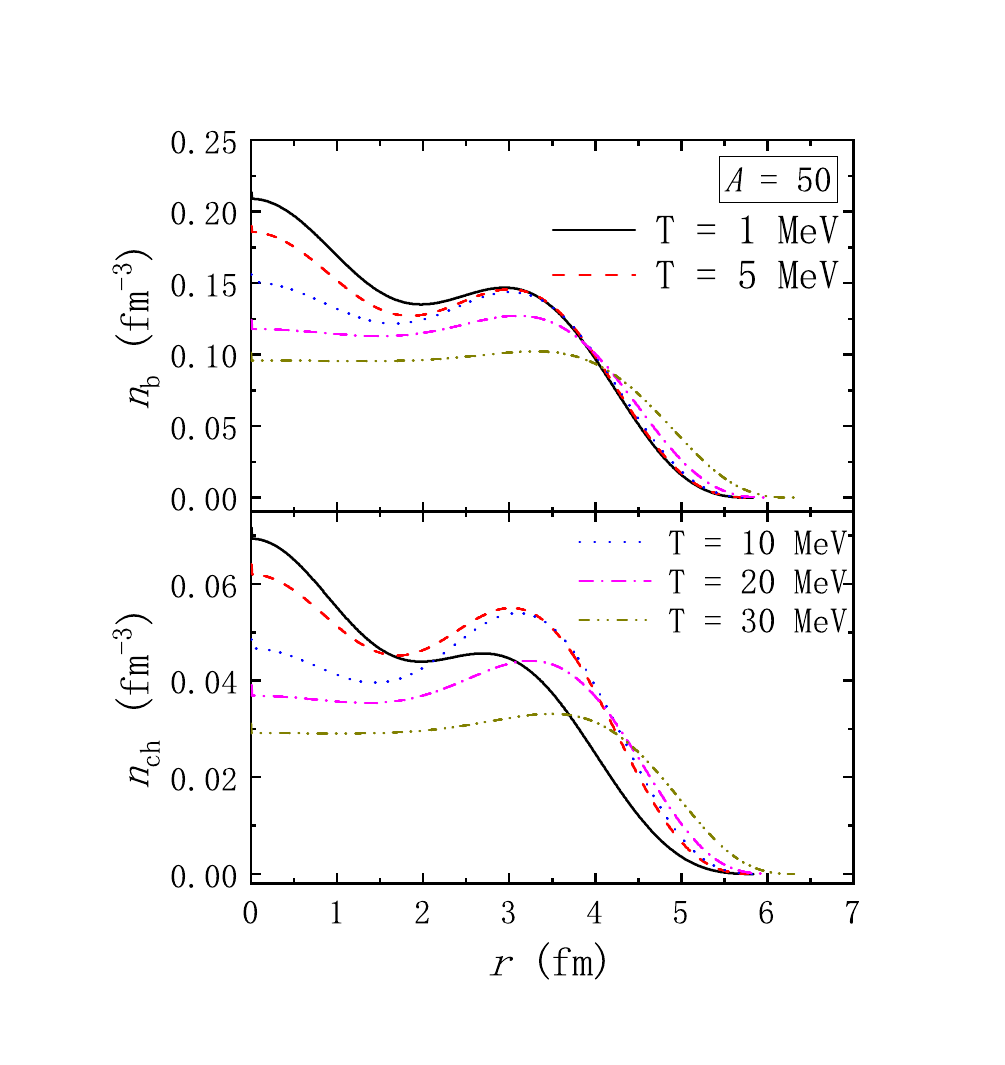}
\caption{\label{Fig:nbch.pdf} The baryon and charge density distributions in a strangelet at various temperatures.}
\end{figure}

Figure~\ref{Fig:nbch.pdf} depicts the charge $n_\mathrm{ch}=(2n_u-n_d-n_s)/3$ and baryon $n_\mathrm{b}=(n_u+n_d+n_s)/3$ density profiles of the strangelets that correspond to Fig.~\ref{Fig:udsDens.pdf}. With an increase in temperature, the baryon number density within the strangelet diminishes and becomes more dispersed in the surface area. The charge density is predominantly positive, which is maximal at the center of the strangelet and decreases with $r$ at low temperatures. With increasing temperature, the variation in charge density becomes smoother, primarily attributed to the alterations in the density profiles of $s$ quarks as depicted in Fig.~\ref{Fig:udsDens.pdf}.

\begin{figure}
\includegraphics[width=\linewidth]{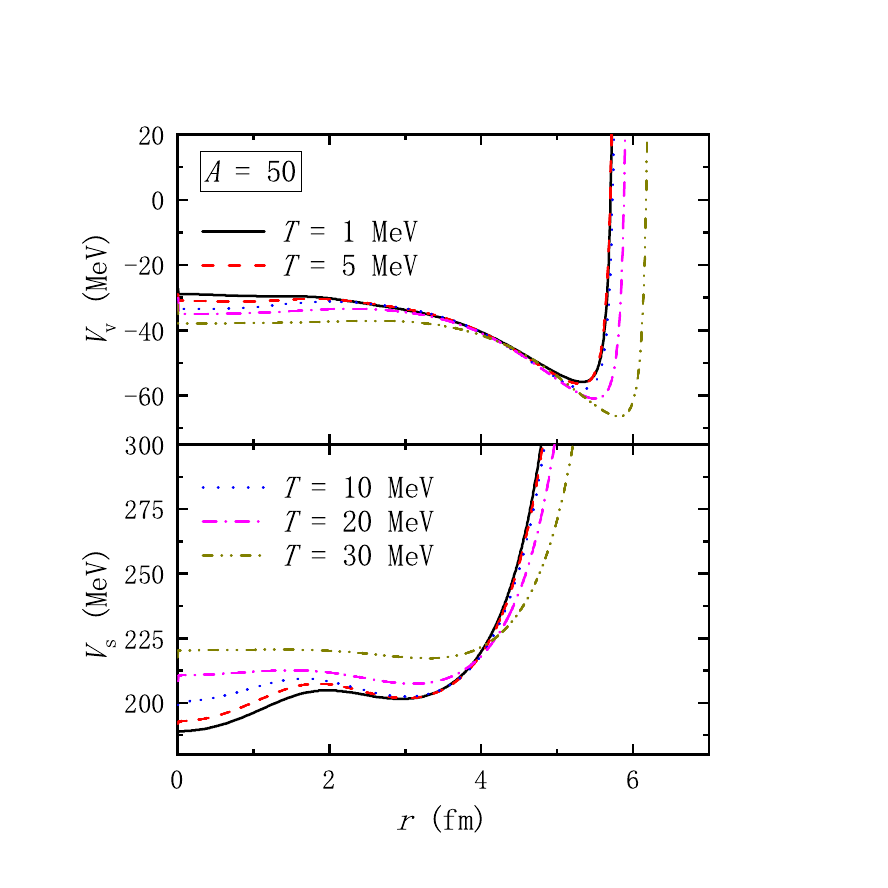}
\caption{\label{Fig:VvVs.pdf} Scalar and vector potentials corresponding to the
density profiles of the strangelets indicated in Figs. \ref{Fig:udsDens.pdf} and \ref{Fig:nbch.pdf}}
\end{figure}

Based on the density profiles indicated in Figs.~\ref{Fig:udsDens.pdf} and \ref{Fig:nbch.pdf}, we obtain the corresponding scalar and vector potentials using Eqs.~(\ref{Eq:Vs}) and (\ref{Eq:Vv}), which are presented in Fig.~\ref{Fig:VvVs.pdf}. By considering quark confinement self-consistently in our mass scaling in Eq.~(\ref{Eq:D}), the mean-field potentials exhibit infinite values near the interface between quark matter and vacuum as illustrated in Fig.~\ref{Fig:VvVs.pdf}. As illustrated by Fig.~\ref{Fig:udsDens.pdf} and \ref{Fig:nbch.pdf}, the densities of strangelets decrease with increasing temperature while their sizes increase. Consequently, the scalar potential depth ($V_S$) becomes shallower at larger temperatures. In contrast, the vector potential ($V_V$) becomes deeper, except for a shift in the location of the infinite potential towards larger strangelets' sizes. At sufficiently high temperatures, there is a significant decrease in densities leading to a rapid expansion of the strangelet's size. Ultimately, the mean fields converge towards the bulk limit, indicating an absence of strangelets at higher temperatures.

\begin{figure}
\includegraphics[width=\linewidth]{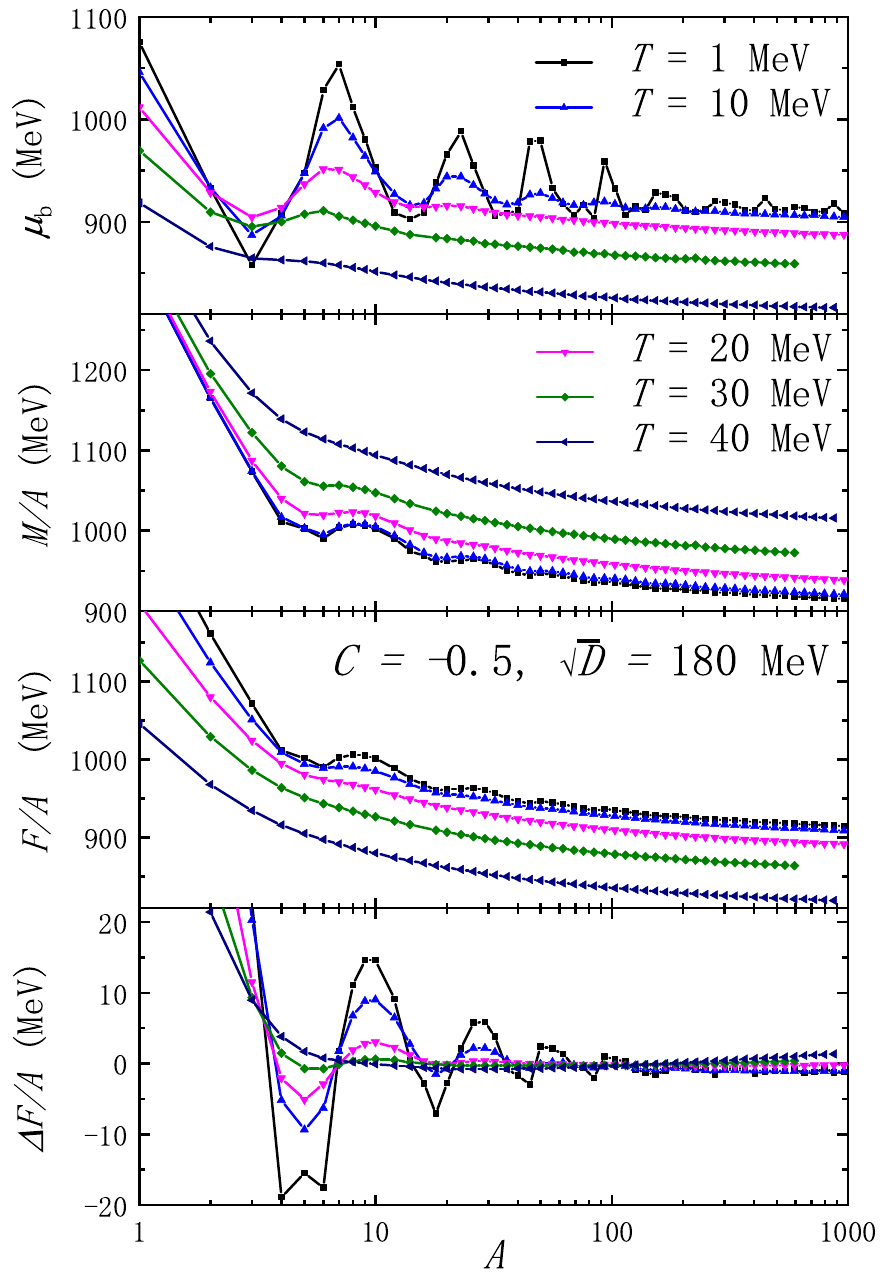}
\caption{\label{Fig:fpaepadf} Chemical potential, energy and free energy per baryon of $\beta$-stable strangelets as functions of baryon number. The deviation in the free energy per baryon $\Delta F/A$ from the liquid-drop formula in Eq.~(\ref{Eq:L-D}) is presented in the lower panel as well.}
\end{figure}

In Fig.~\ref{Fig:fpaepadf}, we present the chemical potential, energy, and free energy per baryon for $\beta$-stable strangelets, plotted against the baryon number $A$ at distinct temperatures, where the parameter set $C$ = -0.5 and $\sqrt{D}$ = 180 MeV is adopted. Based on the Fermi-Dirac distribution outlined in Eq.~(\ref{Eq:F-D}), quarks have a tendency to occupy higher energy states as the temperature rises. Consequently, the energy per baryon increases with temperature $T$, whereas the free energy per baryon and chemical potential exhibit a decreasing trend. Cold strangelets, particularly those with smaller baryon numbers, exhibit fluctuations in their chemical potential, energy, and free energy per baryon as a function of $A$. These fluctuations can be attributed to shell effects~\cite{Xia2018_PRD98-034031}. However, at higher temperatures, the shell effects become less pronounced (i.e., shell damping),  resulting in a smoother variation of $\mu_\mathrm{b}$, $M/A$, and $F/A$ with $A$. Additionally, it is worth noting that at smaller baryon numbers, specifically when $A$ is less than approximately 10, the center-of-mass correction and one-gluon-exchange interactions become significant~\cite{Aerts1978_PRD17-260}. Consequently, the energy and free energy per baryon for $\beta$-stable strangelets are anticipated to decrease. In the extreme scenario where $A$ equals 1 and $T$ is 0, the (free) energy per baryon should align with the mass of nucleons.

The free energy per baryon in Fig.~\ref{Fig:fpaepadf} can be fitted by a liquid-drop type formula~\cite{Oertel2008_PRD77-074015},i.e.,
\begin{equation}
  \frac{F_{LD}}{A} = \frac{F_0}{A} + \frac{\alpha_S}{A^{1/3}} + \frac{\alpha_C}{A^{2/3}} \label{Eq:L-D}
\end{equation}
Here, $F_0/A$ denotes the free energy per baryon in the bulk limit, while  $\alpha_S$ and $\alpha_C$ are the fitted parameters. It's worth noting that during the fitting process, we subtracted the contribution of the Coulomb energy $E_C$ to better highlight the influence of the strong interaction on the interface effects. The deviations from the fitted free energy values, denoted as $\Delta F$ (where $F - E_C - F_{LD}$), are presented in the bottom panel of Fig.~\ref{Fig:fpaepadf}. In this panel, the shell effects can be clearly identified. As the baryon number $A$ increases, the shell corrections to the energy and free energy per baryon eventually become negligible. Similarly, at higher temperatures, quarks begin to occupy a larger number of single-particle states above the Fermi energy, thereby diminishing the fluctuations caused by significant shell gaps, i.e., shell dampening.

 In Fig.~\ref{Fig:fsfzr0.pdf}, the strangeness per baryon, charge-to-mass ratio, and ratio of root-mean-square radius to baryon number can all be derived as
\begin{eqnarray}
  f_s &=& N_s/A,  \label{Eq:fs} \\
  f_z &=& Z/A = (2N_u-N_d-N_s)/3A,   \label{Eq:fz} \\
  r_0 &=& \left(\int 4\pi r^4 n_b \mathrm{d}r\right)^{\frac{1}{2}}/A^{\frac{5}{6}}.   \label{Eq:r0}
\end{eqnarray}
Generally speaking, as the baryon number $A$ increases, the charge-to-mass ratio and the ratio of root-mean-square radius to baryon number tend to decrease, eventually approaching to their bulk values ($f_z = 0$ and $r_0 \approx 0.48/n_{\mathrm{b}}^{1/3}$ with $n_{\mathrm{b}}$ fixed at vanishing pressures). Conversely, the strangeness per baryon increases with A and approaches its bulk value ($f_s \approx 0.7$). At lower temperatures $T$, shell effects induce fluctuations in both the strangeness per baryon and the charge-to-mass ratio due to sequential occupation of lowest energy levels. However, with increasing $A$ or $T$, these fluctuations become less evident. It is noteworthy that the shells exert opposing influences on $f_z$ and $f_s$. Specifically, $f_z$ experiences a sudden drop when the $s$-quark shell is occupied, accompanied by a corresponding surge in $f_s$. This phenomenon also affects the radii of strangelets, as a sudden reduction in $r_0$ is observed when $f_s$ increases. As temperature $T$ rises, the shell effects become less pronounced, resulting in a smoother variation of $f_z$, $f_s$, and $r_0$ with $A$. For the parameter sets $C = -0.5$ and $\sqrt{D}$ = 180 MeV employed in this study, $f_z$ and $f_s$ generally exhibit limited sensitivity to temperature effects beyond shell dampening. Meanwhile, as depicted in Figs.~\ref{Fig:udsDens.pdf} and \ref{Fig:nbch.pdf}, the density profiles of strangelets exhibit a more dilute character as temperature rises, resulting in an increase in $r_0$ as shown in Fig.~\ref{Fig:fsfzr0.pdf}.

\begin{figure}
\includegraphics[width=\linewidth]{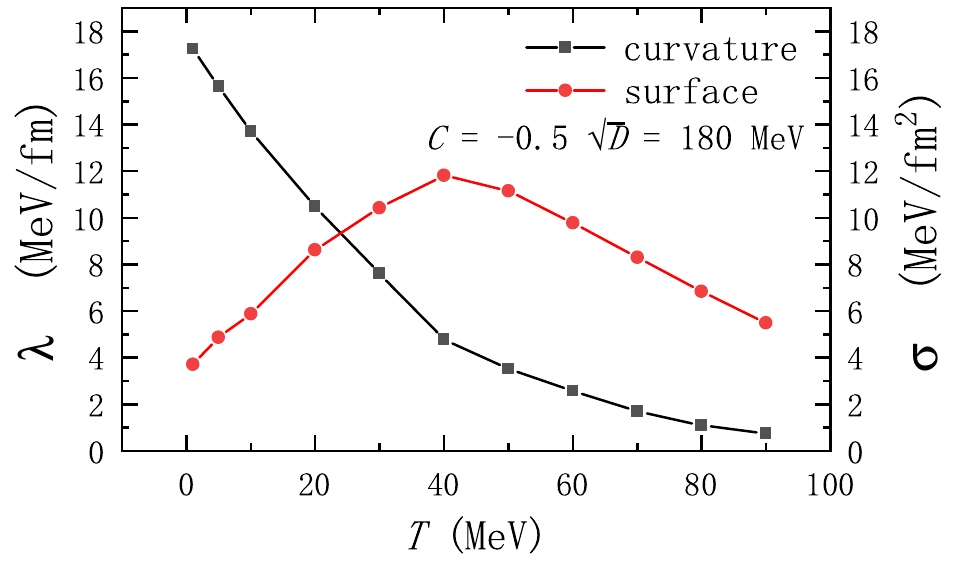}
\caption{\label{Fig:Sur-Cur-T.pdf} The obtained surface tension ¦Ò and curvature term
$\lambda$ of SQM as functions of temperature, which are fixed by Eqs.(~\ref{Eq:Sur}) and (~\ref{Eq:Cur}) }
\end{figure}

The surface tension $\sigma$ and curvature term $\lambda$ can all be obtained with
\begin{eqnarray}
  \sigma &=& \alpha_S \left( \frac{n_0^2}{36\pi} \right)^{\frac{1}{3}},    \label{Eq:Sur} \\
  \lambda &=& \alpha_C \left( \frac{n_0}{384\pi^2} \right)^{\frac{1}{3}}.    \label{Eq:Cur}
\end{eqnarray}
To provide a clearer picture of the temperature-dependent behaviors in the surface tension and curvature term of SQM, in Fig.~\ref{Fig:Sur-Cur-T.pdf} we present the obtained values for $\sigma$ and $\lambda$ plotted against temperature. As denoted by the black squares, the curvature term $\lambda$ decreases steadily with temperature, while the surface tension $\sigma$ rises with $T$ until it attains its peak value at approximately $T \approx 40$ MeV. This is in contrary to our expectations, where the increase in surface tension at temperatures $T$ below 40 MeV is precisely the opposite of what has been typically observed in previous studies with $\sigma$ decreasing with $T$~\cite{Pinto2012_PRC86-025203,Mintz2013_PRD87-036004,Ke2014_PRD89-074041}. The primary reason for this atypical behavior in $\sigma$ can be attributed to shell effects. As temperature $T$ rises, the free energy per baryon subtracted by its bulk value increases rapidly for smaller strangelets due to shell dampening. In this study, our primary focus is on characterizing the properties of all strangelets that may potentially emerge during binary compact star mergers or heavy-ion collisions, so we still use the liquid-drop formula.

\begin{figure}
\includegraphics[width=\linewidth]{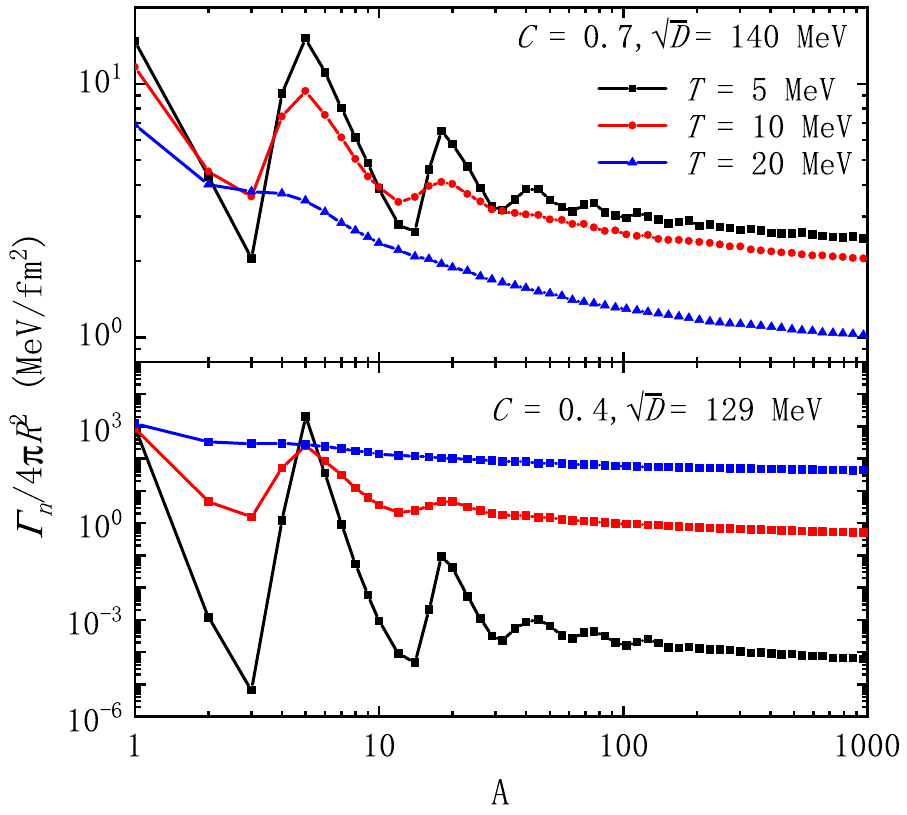}
\caption{\label{Fig:emi2-T.pdf} Neutron emission rates for $\beta$-stable strangelets as functions of baryon number, which are fixed by Eq.~(\ref{Eq:emi}).}
\end{figure}

Utilizing the derived chemical potentials for $\beta$-stable strangelets, the rates of proton and neutron emissions can be determined using Eq.~(\ref{Eq:emi}). To illustrate this, Fig.~\ref{Fig:emi2-T.pdf} shows the neutron emission rates for $\beta$-stable strangelets considering two distinct parameter sets. In particular, compared with nuclear matter, strangelets are stable adopting the parameter set $C$ = -0.5 and $\sqrt{D}$ = 180 MeV, which become unstable for $C$ = 0.7 and $\sqrt{D}$ = 140 MeV. The strangelets obtained with the parameter set $C$ = 0.7 and $\sqrt{D}$ = 140 MeV exhibit instability against neutron emission, characterized by significant decay rates that decrease with temperature. This behavior contrasts sharply with the results obtained using $C$ = -0.5 and $\sqrt{D}$ = 180 MeV, where the neutron emission rates are low and increase with $T$. This variation can be primarily attributed to the evolution of the chemical potential $\mu_{\mathrm{b}}$. Specifically, for unstable strangelets, $\mu_{\mathrm{b}}$ exceeds $m_n$, and $\mu_{\mathrm{b}}$ gradually decreases as temperature rises. As a result of shell effects, the chemical potential experiences fluctuations that are dependent on the baryon number $A$. Consequently, the emission rates also undergo corresponding variations. This gives rise to the existence of extremely stable strangelets, which might serve as key waiting points during the decay process of larger strangelets. These stable strangelets are likely to persist in binary compact star mergers or heavy-ion collisions. However, at higher temperatures, these fluctuations in nucleon emission rates are suppressed due to shell dampening, while the neutron and proton emission rates per unit area may decrease significantly, attributed primarily to the rapid decline in the chemical potential.

\section{\label{sec:summary}Summary}
Employing equivparticle model with density-dependent quark masses, we investigate the interface effects and properties of $\beta$-stable strangelets with various parameter sets, accounting for both linear confinement and leading-order perturbative interactions. Based on the mean-field framework, we study various strangelets' characteristics, such as the density profiles of $u$, $d$, and $s$ quarks, the distribution of baryon and charge densities, the free and total energy per baryon, the chemical potential, the strangeness content per baryon, the charge-to-mass ratio, as well as the nucleon emission rates. As temperature rises, the shell corrections eventually lose significance as quarks occupy higher energy states, a phenomenon known as shell dampening. Consequently, contrary to the expected trend, the surface tension, derived from a liquid-drop formula fitted to the free energies of strangelets, increases with temperature until it peaks at $T \approx 20-40$ MeV, where shell corrections vanish. In contrast, the curvature term decreases monotonically with temperature. Using the derived properties of strangelets, we can determine the neutron and proton emission rates by applying standard formulae for nuclear reaction rates from the theory of nucleosynthesis. These calculations incorporate the equilibrium densities of the external nucleon gas and the capture cross sections of strangelets. The emission rates of stable strangelets generally increase with temperature, while unstable strangelets exhibit monotonically decreasing emission rates with temperature. These emission rates are intimately linked to the chemical potential of strangelets, which experiences fluctuations due to shell corrections. Consequently, certain strangelets exhibit remarkable stability against neutron or proton emissions.

% \begin{acknowledgments}
\section{Acknowledgments}
This work was supported by the National Natural Science Foundation of China (Nos.~12275234 and 12342027), the Strategic Priority Research Program of the Chinese Academy of Sciences (No.~XDB0550300),
and the National SKA Program of China (No.~2020SKA0120300 and No.~2020SKA0120100).
% \end{acknowledgments}

%%
%% reference here
%%
\bibliography{strange_quark}

\end{document}